\documentclass[10pt,A4paper,conference]{IEEEtran}
\IEEEoverridecommandlockouts
\usepackage{cite}
\usepackage{amsmath,amssymb,amsfonts}
\usepackage{graphicx}
\usepackage{textcomp}
\usepackage{xcolor}
\usepackage{tabularx}
\usepackage{color, colortbl}

\definecolor{lime}{rgb}{0.88,2,10}
\usepackage[left=.55in,
            right=.55in,
            top=.69in,
            bottom=0.9in]{geometry}

\renewcommand{\baselinestretch}{.945}
\usepackage{subcaption}
\usepackage{wrapfig}
\usepackage[linesnumbered,ruled,vlined]{algorithm2e}
\usepackage{xpatch}
\usepackage{url}
\usepackage{multirow}
\usepackage{multicol}
\usepackage{wrapfig}

\usepackage[export]{adjustbox}
\usepackage{nomencl}
\usepackage{siunitx}
\usepackage{blindtext}
\usepackage[T1]{fontenc}
\usepackage[spaces,hyphens]{xurl}
\usepackage{float}

\newcommand*{\Resize}[2]{\resizebox{#1}{!}{$#2$}}%

\SetAlFnt{\small}

\SetKwComment{Comment}{$\triangleright$\ }{}

\usepackage[utf8]{inputenc}
\usepackage{enumitem}

\SetKwInput{kwInit}{Initialize}

\def\BibTeX{{\rm B\kern-.05em{\sc i\kern-.025em b}\kern-.08em
    T\kern-.1667em\lower.7ex\hbox{E}\kern-.125emX}}

\usepackage[numbers,sort&compress]{natbib}
\usepackage[font={small}]{caption} 

\newcommand{\fref}[1]{Fig.~\ref{#1}}

\usepackage[misc]{ifsym}

\usepackage{fancyhdr}
\fancypagestyle{mystyle}{
    \chead{\small 2023 IEEE International Conference On Metrology for eXtended Reality, Artificial Intelligence, and Neural Engineering (MetroXRAINE)}
}

\title{A Brain-Computer Interface Augmented Reality Framework with Auto-Adaptive SSVEP Recognition}
\author{\IEEEauthorblockN{
Yasmine Mustafa\IEEEauthorrefmark{1}\IEEEauthorrefmark{4}, Mohamed Elmahallawy\IEEEauthorrefmark{1}, Tie Luo\IEEEauthorrefmark{1},   Seif Eldawlatly \IEEEauthorrefmark{2}\IEEEauthorrefmark{3}}  
      \IEEEauthorblockA{%
  \IEEEauthorrefmark{1}Computer Science Department, Missouri University of Science and Technology, MO, USA}
        \IEEEauthorblockA{%
        \IEEEauthorrefmark{2}Computer and Systems Engineering Department, Ain Shams University, Egypt}

  \IEEEauthorblockA{%
    \IEEEauthorrefmark{3}Computer Science and Engineering Department, The American University in Cairo, Egypt}
      \IEEEauthorblockA{%
    \IEEEauthorrefmark{4}Digital Media Engineering and Technology Department, German University in Cairo, Egypt}

  Emails:  \{yam64, tluo, meqxk\}@mst.edu, seldawlatly@eng.asu.edu.eg
\vspace{-0.2in}
}

\begin{document}

\maketitle
\thispagestyle{mystyle}

\begin{abstract}
Brain-Computer Interface (BCI) initially gained attention for developing applications that aid physically impaired individuals. Recently, the idea of integrating BCI with Augmented Reality (AR) emerged, which uses BCI not only to enhance the quality of life for individuals with disabilities but also to develop mainstream applications for healthy users.  One commonly used BCI signal pattern is the Steady-state Visually-evoked Potential (SSVEP), which captures the brain's response to flickering visual stimuli. SSVEP-based BCI-AR applications enable users to express their needs/wants by simply looking at corresponding command options. However, individuals are different in brain signals and thus require per-subject SSVEP recognition. Moreover, muscle movements and eye blinks interfere with brain signals, and thus subjects are required to remain still during BCI experiments, which limits AR engagement. In this paper, we (1) propose a simple adaptive ensemble classification system that handles the inter-subject variability, (2) present a simple BCI-AR framework that supports the development of a wide range of SSVEP-based BCI-AR applications, and (3) evaluate the performance of our ensemble algorithm in an SSVEP-based BCI-AR application with head rotations which has demonstrated robustness to the movement interference. Our testing on multiple subjects achieved a mean accuracy of 80\% on a PC and 77\% using the HoloLens AR headset, both of which surpass previous studies that incorporate individual classifiers and head movements. In addition, our visual stimulation time is 5 seconds which is relatively short. The statistically significant results show that our ensemble classification approach outperforms individual classifiers in SSVEP-based BCIs.
\end{abstract}

\section{Introduction}
Brain-computer Interface (BCI) enables the human brain to interact with an external device by sending and receiving just brain signals, allowing people to control computers or other devices with no stimulation of muscles. It uses electrodes to capture brain activity and translates it to commands that can be sent to external devices to take desired actions \cite{pal2022brain}.  In the past, investigations in the field of brain activity were limited to medical cases and studying neurological disorders in clinics, but recently, computing and biosensing innovations have significantly improved the outlook for BCI applications, and many studies have appeared that create assistive tools for people with disability to regain control and communication abilities \cite{pal2022brain}, such as controlling a prosthetic arm \cite{mcfarland2011brain}.

More recently, BCI has started to evolve only as an aid for physically impaired individuals to games and mainstream applications for normal people \cite{tezza2020brain}. One direction for developing mainstream BCI applications is integrating BCI with Augmented Reality (AR). Unlike Virtual Reality (VR) which completely isolates the user from the actual environment, AR provides an environment that shows both the virtual contents and the actual environment. By doing so, it offers users a middle ground between being immersed in the application and being present in the real world. 

The idea of merging BCI with Extended Reality (XR), such as AR and VR, was introduced by Jantz et al. \cite{jantz2017brain} to address the limitations of XR controls such as hand gestures and vocal commands.  For example, hand gestures break the user's immersion in AR and VR, limit users to objects at arm length, and have safety concerns of hitting objects in the physical world. Voice controls function poorly in loud environments and can be socially embarrassing. Thus, the integration of BCI and AR opens up new opportunities in gaming, enterprise, and medical applications. Hence, integrating BCI (instead of external controls) and AR as an emerging paradigm is more promising which both enhances user experience and opens up new application opportunities. Moreover, combining BCI and AR 

The most popular BCI is the electroencephalogram (EEG) based BCI because of its safety and reasonable temporal resolution. One of the control signals used in BCI based on EEG systems is the Steady-State Visually-Evoked Potentials (SSVEPs). SSVEP is a resonance phenomenon that can be observed through electrodes placed on the scalp (specifically over the occipital and parietal lobes of the brain) when a subject looks at a light source (stimulus) flickering at a constant frequency \cite{icscan2018steady}. When a person gazes at a specific stimulus, SSVEP signals are generated in the brain as a natural response to the stimulus with the same frequency or multiples of the frequency of that stimulus. In EEG recordings, the spectral power of the SSVEP responses is compared and analyzed for different purposes. One popular usage is to relate each EEG tag to a specific control command to design an application, where such an SSVEP-based BCI application allows the user to access a variety of control commands simply by focusing her attention on them.

One of the first studies that introduced BCI-AR into mainstream applications is Si-Mohammed et al. \cite{si2018towards}. They focused on examining how head movements and how the layout of the designed visual stimuli affect performance. In the online experiments, mean accuracy, when head movements were incorporated, dropped from 78\% to 41\%, indicating the need for more robust systems. In addition, the experiments also showed that the performance depended on the match between the layout design and user preference, indicating the issue of subject dependency when it comes to brain signals.  

Addressing subject-dependent brain signals, Yue et al. developed an attention-aware system \cite{vortmann2020attention} which tries to limit the virtual content displayed to a user in order to reduce distractions caused by surroundings to her, thereby improving the performance of SSVEP recognition by aligning better with the user's brain signals.  Although considering the attentional state can serve as a useful guideline for adjusting visual content, it alone is insufficient to address all the complexities of subject-dependent brain signals. In fact, relying solely on attentional state information yielded a mean accuracy of only 65.6\% \cite{vortmann2020attention}.

A more recent study was conducted by Zhang et al., where they focused on controlling a robot's movement in a maze \cite{zhang2022humanoid}. Zhang et al. used AR as a portable visual stimulation device, which means that the user does not need to switch their focus back and forth between the environment and the stimulation device. Results showed that all subjects successfully completed the robot walking task, demonstrating the feasibility of the AR-SSVEP-NAO system. Ke et al. also showed promising results in controlling a robotic arm in a high-speed online SSVEP-based BCI-AR environment and discussed the limitations that can be imposed by head-mounted displays such as the HoloLens as they can provide less stable frame rates than the computer screens \cite{ke2020online}. 

The mentioned work shows the feasibility of integrating AR and BCI and encourages the development of more applications and ideas for hands-free and voice-free user interfaces \cite{vortmann2020attention}. These studies inspired us to extend the exploration of AR-BCI environments and overcome common design challenges: 1) the artifacts that can be introduced to the EEG signal through blinking eyes and muscle movements require the user to stay still throughout the entire experiment, and 2) the uniqueness of the EEG signal of each individual creates the variation of performance among subjects. Meeting these challenges necessitates an approach that can handle unique brain signals, while also being robust enough to accommodate essential movements in an AR environment, like head rotations, that introduce artifacts during the process.

In light of these challenges, this paper makes the following contributions: 

\begin{itemize}[leftmargin=*]
    \item We propose a BCI-AR framework that is robust to artifacts and easy to implement for creating SSVEP-based BCI-AR applications. 
     
    \item To allow an immersive AR user experience, we allow head movements which was difficult in previous studies due to the sensitivity of brain signals to various factors such as external distractions, user's inattention and muscle movements.
    
    \item We utilize an ensemble SSVEP recognition system that auto-adapts to different subjects and movements and show that it is better than individual classification models.
    
    \item Our experiments of using the HoloLens and PC with BCI demonstrate that the HoloLens provides accuracy close to that provided by the PC, which means that the HoloLens is on par with the PC, the conventional device used with BCI. Our experiments show promising results that are superior to previous studies that allowed head movements in BCI-AR systems, as well as better accuracy across different subjects, indicating good generalizability.
   
\end{itemize}

\section{Framework}

In this section, we begin by presenting an overview of the proposed system, followed by a description of its application. Then we highlight the key specifications of the SSVEP interface to provide a better understanding of its importance.
\subsection{System Overview}
\begin{figure}[!t]
  \centering
  \includegraphics[width=0.8\linewidth]{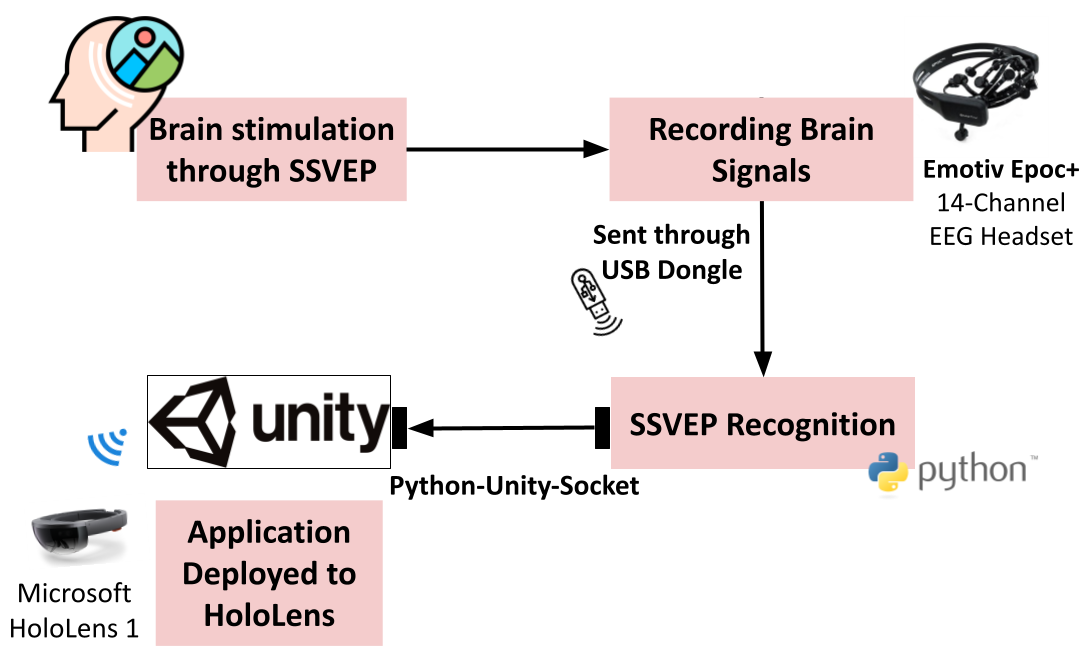}
     
  \caption{Overview of the full pipeline between BCI implemented in Python and AR implemented in Unity.}
  \label{Images/ConnectionOverview.pdf}
\end{figure}

The proposed system comprises two essential hardware components: the first-generation Microsoft HoloLens and the Emotiv Epoc EEG headset.  The software implementation involves an AR interface, EEG signal recording, preprocessing, classification, and the interface between the AR and BCI components. \fref{Images/ConnectionOverview.pdf} shows an overview of the framework where the HoloLens AR headset displays buttons as visual stimuli, while the Emotiv headset records and transmits EEG signals for processing in Python. The recorded data is then read, preprocessed, and classified using our ensemble machine-learning model. The classification results are communicated from Python to Unity via a Python-Unity-Socket \cite{Python-Unity}. Unity, in turn, presents relevant feedback to the user on the HoloLens in response to each input.

\subsection{Application}
The application implemented in this work is developed using Unity on top of an architecture of an open-source HoloLens project  \cite{HololensVocalCommands}. This architecture originally had three vocal commands that correspond to three actions: \textit{Create Cube}, \textit{Create Sphere}, and \textit{Delete All}. 

\begin{figure}[ht]
  \centering
  \includegraphics[width=0.9\linewidth,height=4.9cm]{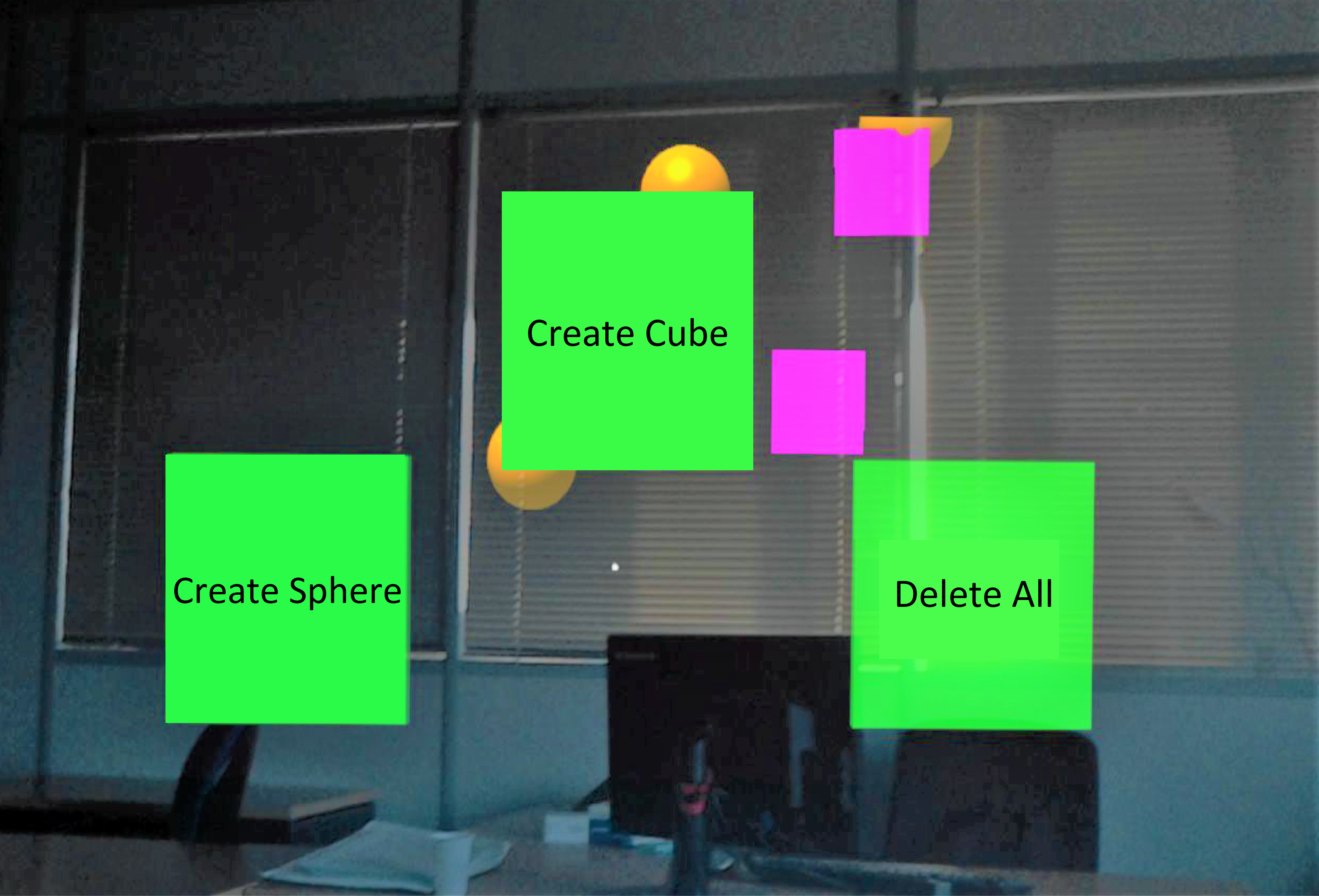}
  \caption{Our application deployed on the HoLoLens. The three green buttons correspond to 3 different commands to be selected using BCI input.}
  \label{Images/3_buttons.pdf}
\end{figure}

Since we want to prove that we can replace the vocal commands with SSVEP-based BCI input, we added three flickering buttons (visual stimulation) to the scene that correspond to the same three commands. If \textit{Create Cube} button is selected (via the SSVEP-based BCI), a cube is created in space where the cursor of the HoloLens is pointed at. Similarly, selecting the \textit{Create Sphere} button results in the creation of a sphere at the cursor location. Lastly, choosing the \textit{Delete All} button removes any previously created shapes. \fref{Images/3_buttons.pdf} demonstrates the design of the BCI-AR application in which the three green buttons are the flickering buttons that represent the visual stimulation, and the shapes created in space are examples of the feedback provided to the user upon her/his button gaze selection. 

\subsection{SSVEP Interface Specifications}

Flickering frequencies and colors of the buttons are selected based on the study done by Duart et al. \cite{duart2021evaluating}. The frequencies assigned to each button are as follows: \textit{Create Cube} flickers at a frequency of 12 Hz, \textit{Create Sphere} at 8.57 Hz, and \textit{Delete All} at 10 Hz. Duart et al. studied the effect of colors on the signal-to-noise ratio (SNR) of the EEG signal. They used the red, white, and green colors and frequencies 5 Hz (low frequency), 12 Hz (medium frequency), and 30 Hz (high frequency) in their study.  For low frequencies, the green color was found to show the best results. In our system, given that we use low flickering frequencies, the green color was selected for the buttons.

\section{Methods}
 
In this section, we provide comprehensive descriptions of both the hardware and software components used in this study. The software encompasses various aspects, including data specifications, preprocessing techniques, and classification methods. 

\subsection{Hardware}

For EEG recording, we used the Emotiv Epoc+, a 14-channel EEG headset with electrodes positioned at $AF3,$ $F7, F3, FC5, T7, P7, O1, O2, P8, T8, FC6, F4, F8, \text{and} AF4$. These electrodes are saline-based and do not require the use of sticky gel, ensuring a more comfortable experience for the user.

\begin{figure}[t!]
  
  \centering
 {\includegraphics[width=0.25\linewidth]{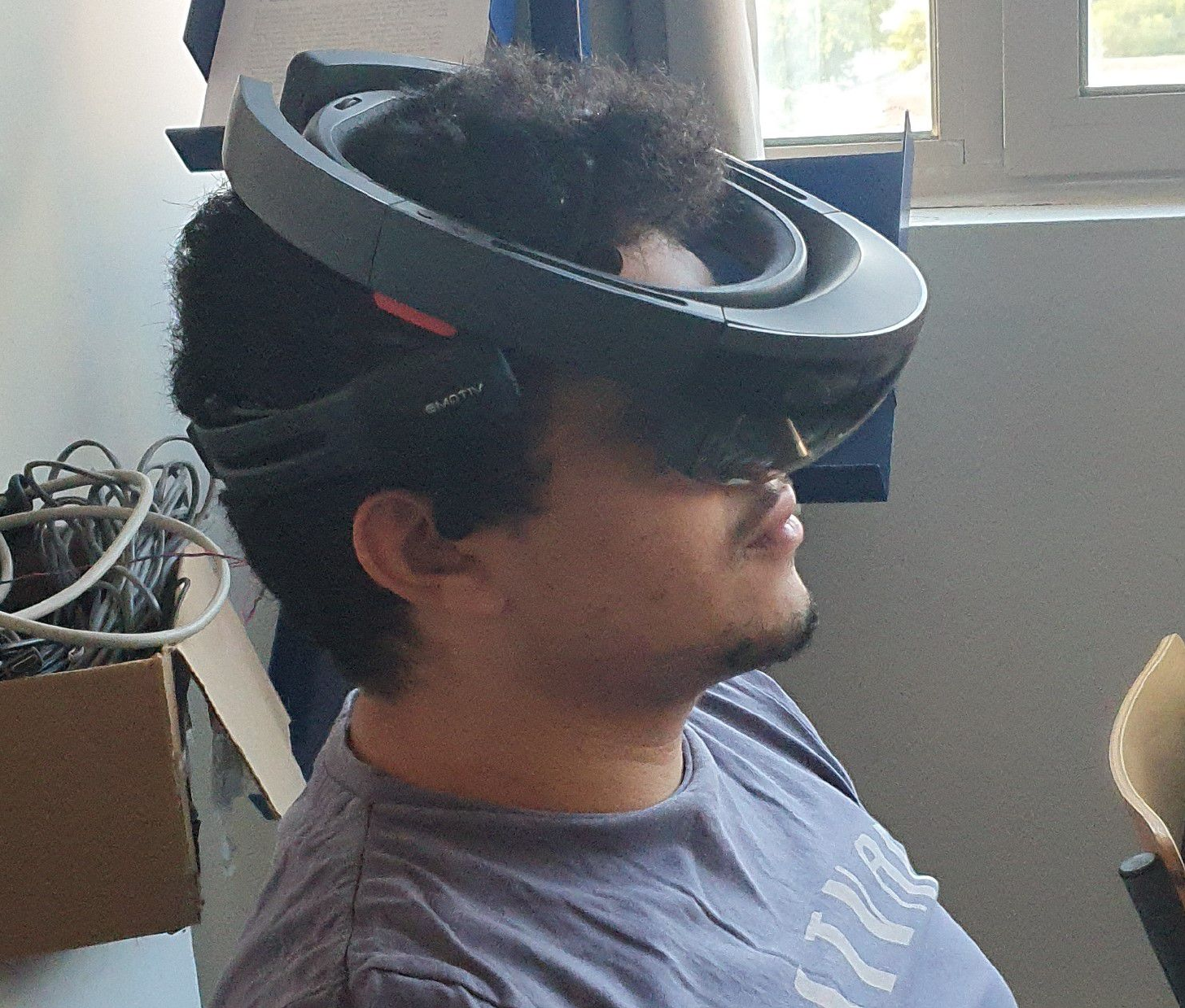}\label{fig:s1}}
  \hfill
  {\includegraphics[width=0.245\linewidth]{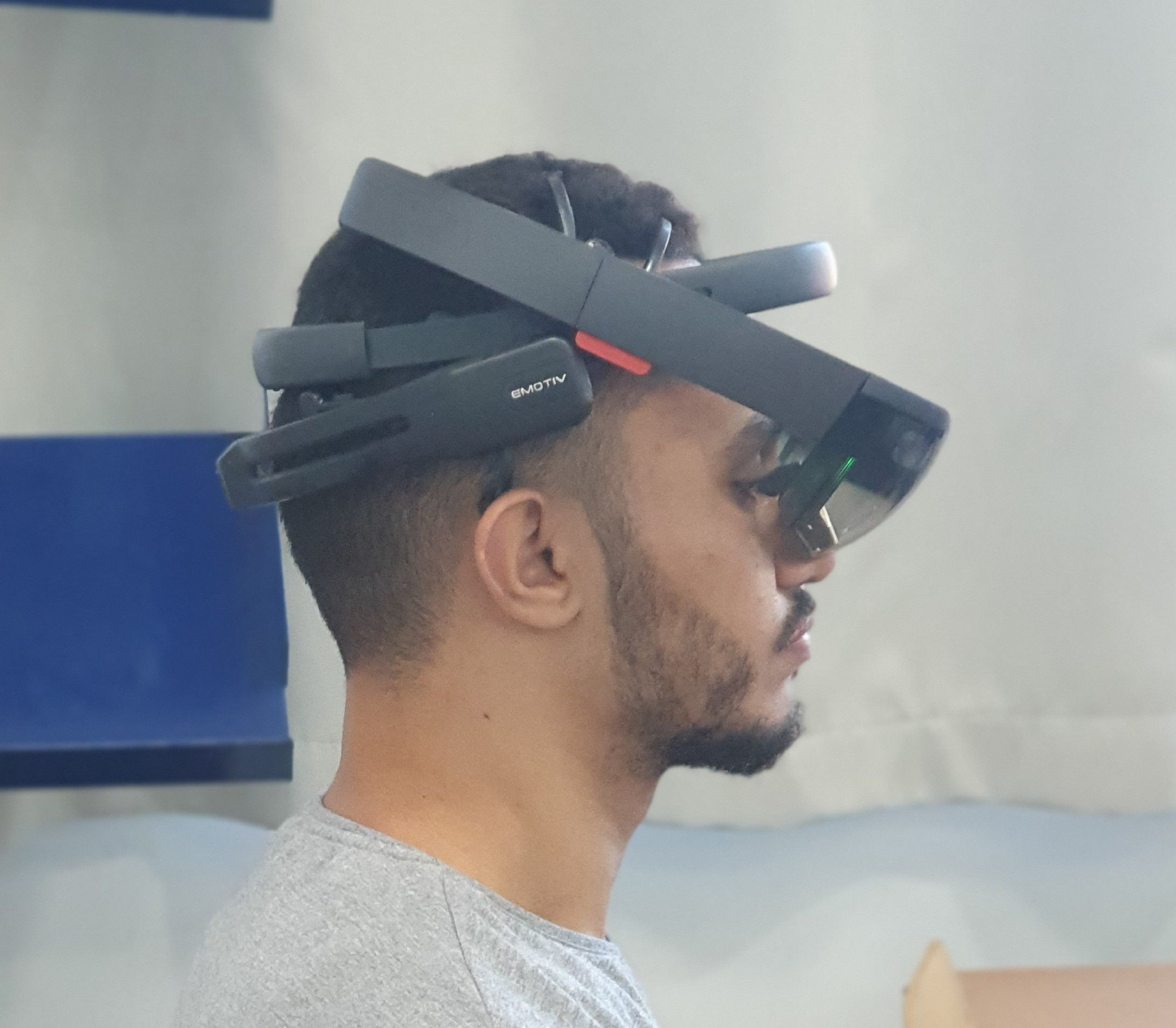}\label{fig:s2}}
    \hfill
 {\includegraphics[width=0.225\linewidth]{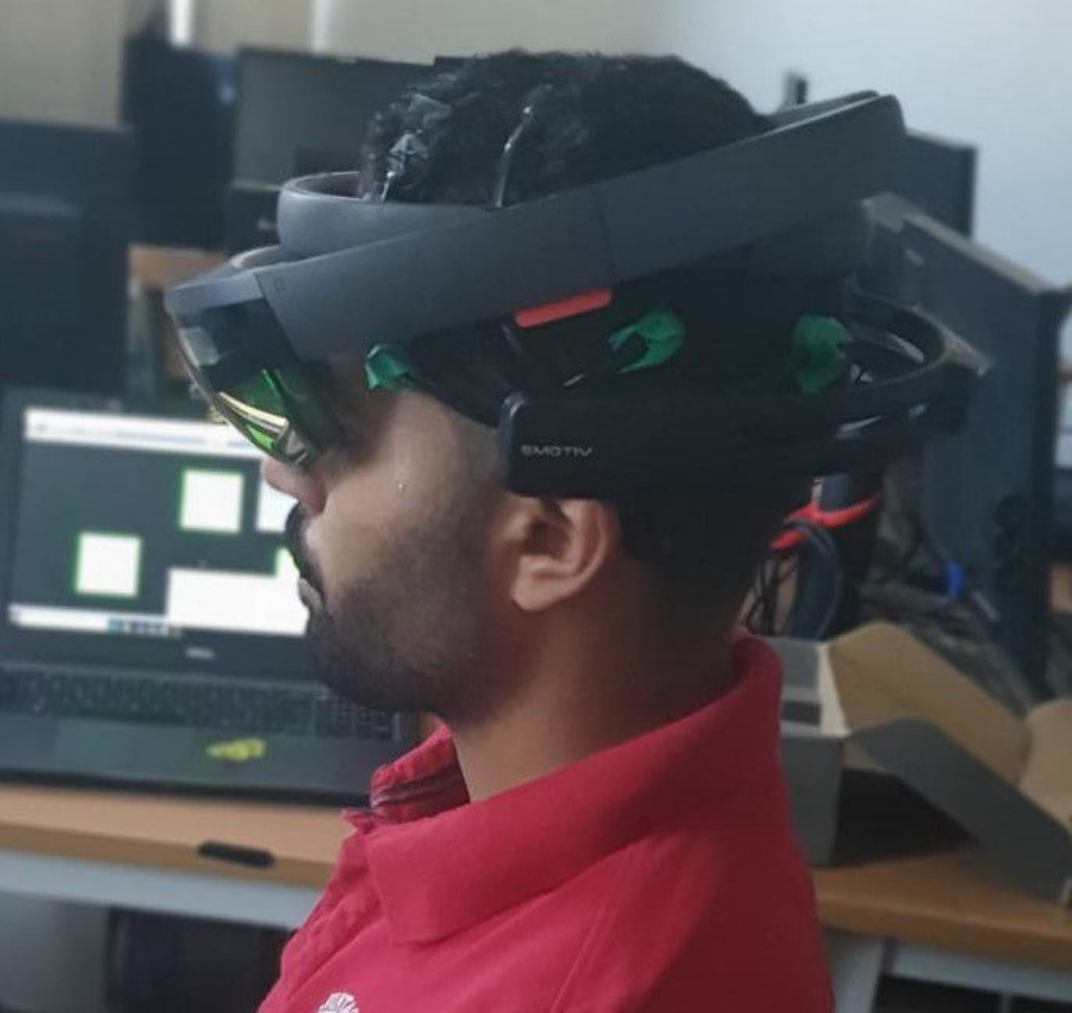}\label{fig:s3}}
  \hfill
  {\includegraphics[width=0.24\linewidth,height=1.9cm]{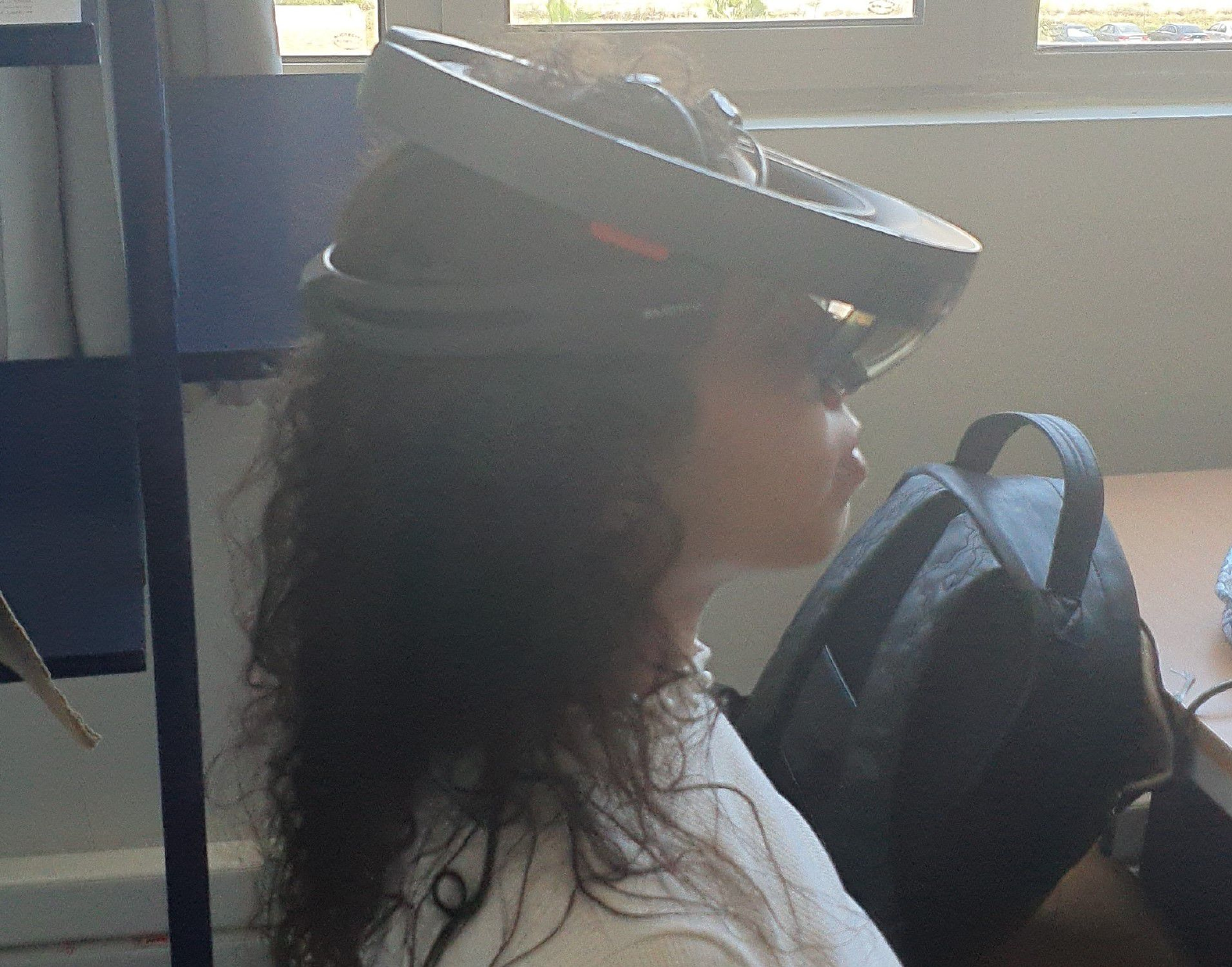}\label{fig:s4}}
  \caption{Subjects wearing the Emotiv Headset and the HoloLens.}
   \label{Images/subject}
\end{figure}

In the context of AR, we utilized the first generation of Microsoft HoloLens. This AR device enables the display of 3D content seamlessly integrated with the real world surrounding the user. To enhance user comfort, the HoloLens is equipped with nose pads, overhead straps that reduce its weight burden, and a rotating headband that can be adjusted to fit the shape and size of each user. For visual reference, Figure \ref{Images/subject} showcases the participation of different subjects in this study, all wearing both headsets simultaneously.

\subsection{Software}

\subsubsection{Data Recording Specifications}
Each of the three buttons in the application flicker with its specified frequency for 5 seconds followed by a resting period of 5 seconds, creating what is referred to as a \textit{trial} that has a total duration of 10 seconds. A \textit{session} is the term used when a number of trials are conducted consecutively. For each subject, 5 sessions are conducted. Each session is divided into two parts: Session \textit{a} contains 12 trials, and session \textit{b} contains 13 trials, creating a total of 25 trials for each session. There is a 30-second resting period between sessions \textit{a} and \textit{b}. A synchronization clock organizes the start and end of both the data recording of the Emotiv headset and the flickering of the buttons such that the Emotiv does not record the resting periods and records only during the flickering part of the trial. For each trial recorded, there are 14 columns that correspond to the 14 channels and 1285 rows for each channel. The sampling rate is 257 Hz. 

\subsubsection{Preprocessing and SSVEP Recognition}

First, common average reference (CAR) filter is applied. The CAR filter is a spatial filter that removes the averaged brain activity across all electrodes for a specific time instant. This helps to remove the common features across all electrodes, hence reducing noise \cite{bertrand1985theoretical}. The formula used to compute the CAR filter takes the following form

\begin{equation}
   \Resize{6cm}{  {U_i(t)}^{CAR} =  {U_i(t)}^{ER} - \frac{1}{n} \sum_{j=1}^n {U_j(t)}^{ER}}
    \label{eq_Car}
 \end{equation}
 where $ {U_i(t)}^{ER}$ is the potential difference between the $i^{th}$
electrode and the ear reference, and $n$ is the 
number of electrodes in the headset. For every electrode $i$, the average is subtracted from the potential difference, resulting in  $ {U_i(t)}^{CAR}$.

Next, Fast Fourier Transform (FFT) is applied to the signal recorded on each channel (data recorded from an electrode). Next, only the flickering frequencies and their harmonics are extracted from the resulting power spectrum. The peak is not necessarily visible at the frequency itself. This may be due to the head rotations and eye blinking as muscle movements are known in the literature to introduce artifacts in the signal \cite{bakardjian2010optimization}, but it's more likely to be due to the unstable frame rates in Microsoft Hololens \cite{arpaia2022performance}. To compensate for this shift in peak frequencies, a window of size 1 Hz, ranging from 0.5 Hz before the frequency to 0.5 Hz after the frequency, is taken at each frequency and its harmonics. 

Next, Principal Component Analysis (PCA) is applied to compress important features into fewer columns and decrease the running time \cite{rabby2021epileptic}. Data is then normalized using the Z-score method \cite{al2006normalization}.

Online experiments are done using channels $O1$ and $O2$ only as these are the electrodes covering the visual cortex area where SSVEP is most prominent. However, to analyze and compare the results, preprocessing is done offline using the concatenation of all the channels first. 
\subsubsection{Classification Techniques}

Mainly the two machine learning classifiers used in the experiments are Support Vector Machine (SVM) and Random Forests (RF). SVM has been widely used by researchers to classify EEG data. In previous studies, it showed better performance than neural networks \cite{sacca2018classification}. SVM performed best in this work with linear and polynomial kernels. RF as well showed promising results in recent studies that classified EEG data \cite{yue2021exploring}. 

The classification of the offline experiments involved splitting the recorded data into two sets: 80\% for training and 20\% for testing, ensuring that each set maintained equal proportions of each class. As our approach is subject-based, we implemented \textit{subject-wise stratified partitioning}. Subject-wise stratified partitioning ensures that the data from each individual subject are kept together while maintaining data of each class within the training and testing sets, allowing the model to learn subject-specific patterns.

The ensemble model consists of four varieties of each classifier: PCA and CAR filter, CAR filter only, PCA only, and without PCA or CAR filter. This results in eight variations: four for the SVM classifier and four for the RF classifier.

\begin{figure}[t!]
  \centering
  \includegraphics[width=\linewidth]{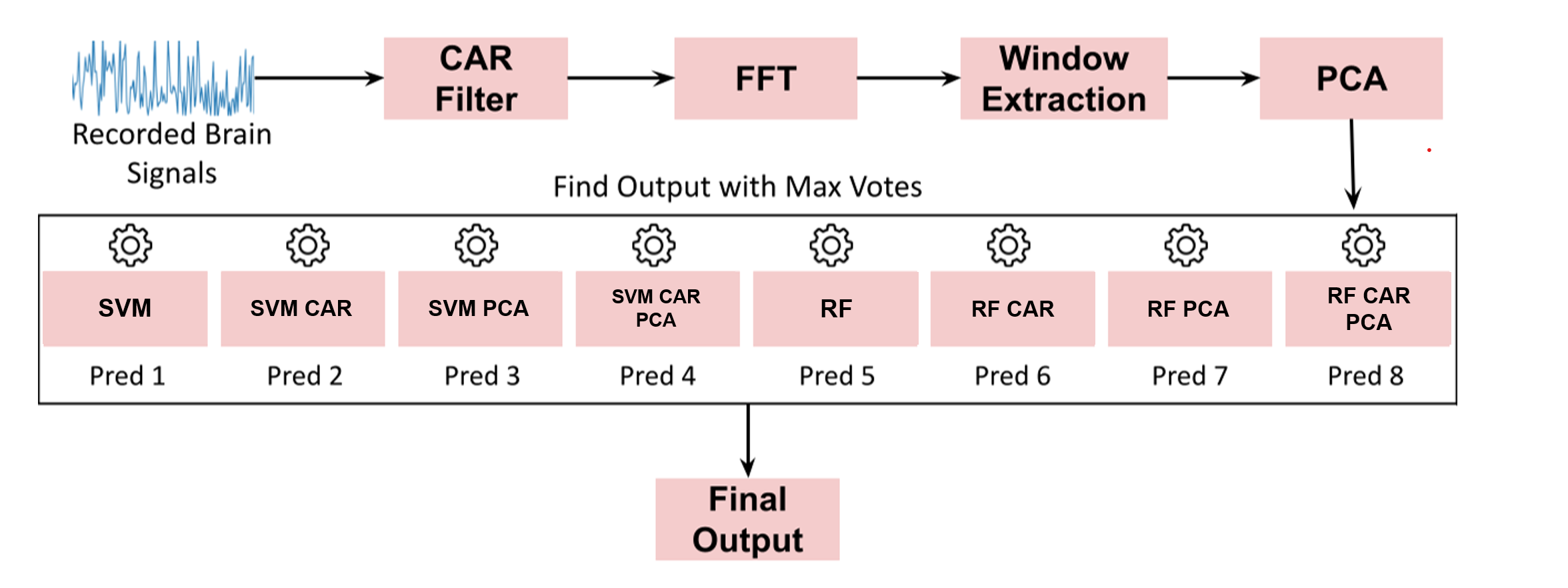}
  \caption{Procedure of the online preprocessing and classification per trial.}
  \label{Images/online_system.png}
\end{figure}

For every trial, the predictions of all models are considered and the label with the highest weighted number of votes is the output of the ensemble model. The vote of every label is multiplied by the training accuracy of the model that predicted that label, creating a weighted ensemble classifier \cite{hussain2021ensemble}. Weighted votes give more weight to classifiers that are more likely to give better performance and less weight to classifiers that are less likely to give good results. Thus, classifiers with worse performance are prevented from decreasing the overall ensemble accuracy. The final accuracy is calculated based on the output that got a maximum number of votes. \fref{Images/online_system.png} gives an overview of the online system. CAR filter and FFT both demonstrated high performance compared to other preprocessing methods in the literature \cite{zerafa2018train,kolodziej2016comparison}.

\subsection{Evaluation Metrics}
The performance of most BCI systems is evaluated using accuracy and Information Transfer Rate (ITR) \cite{hsu2020extraction}. Accuracy is the percentage of correctness in predicting the output of the user gaze. A correct output happens when the output label is that of the command the user gazed at. Accuracy is defined as the total number of correct predictions divided by the total number of predictions. On the other hand, ITR does not only consider the number of correct predictions but also relates the accuracy with the number of classes and stimulation time. ITR can be defined as follows where $B$ is the information transferred in bits per trial, $N$ is the number of commands, and $P$ is the classification accuracy \cite{wolpaw2002brain}.
\begin{equation}
 \Resize{7.8cm}  { B\Big({Bit \over Trial}\Big) = log_2 N + P \times log_2P+(1-P) \times log_2 \Big({1-P \over N-1}\Big)}
\end{equation}
To calculate the information transferred in bits per minute, the result of the above equation, $B$, is multiplied by the average classification time in minutes, $Q$ \cite{wolpaw2002brain}.
\begin{equation}
  \Resize{3.5cm}   {ITR\Big({Bit \over Min}\Big) = B \times Q}
\end{equation}

\section{Results}
We consider four different experiments to evaluate our proposed method: i) offline experiments on the PC, ii) offline experiments on the HoloLens, iii) online experiments on the PC, and iv) online experiments on the HoloLens. Multiple experiments with different setups (design and layout) were performed to identify the setup that gives the best online performance. A total of six volunteers took part in the experiments, with three assigned to the PC experiments (Subjects S1, S2, and S3) and the remaining three assigned to the HoloLens experiments (Subjects S4, S5, and S6). The age range of the participants was between 18 and 26 years. In order to ensure a fair and unbiased assessment, there was no preference given to specific genders or hair types (in terms of thickness or length), aiming to maintain an equitable mean accuracy. Volunteers from all backgrounds were welcomed and actively encouraged to participate in the study.

\textit{Offline experiments} were used to describe the data analysis conducted after recording the data, where no real-time feedback is provided to the user.  The experimental sessions consisted of five individual sessions, each further divided into session \textit{a} and session \textit{b}. In session \textit{a}, the user selects buttons based on a predetermined sequence, and a different sequence is used in session \textit{b}. For instance, a sequence can be 2, 0, 1, 2, 0, 2, 1, 0, 1, 0, 1, 2, where 0 corresponds to "create cube", 1 corresponds to "delete all", and 2 corresponds to "create sphere". The user is required to direct their gaze to the respective buttons as indicated by the appearance of a red dot during each trial. A resting period of 5 seconds is provided between trials to allow the user to shift their focus to the next button.  
Following the offline experiments, the \textit{online experiments} involve providing feedback to the user based on their gaze selection, while also allowing them to freely move their head to navigate the environment and determine the desired location for creating shapes using the head cursor. The classification model used in the online experiments is pre-trained during the offline experiments.

\subsection{Experiments using Personal Computer (PC)}
\subsubsection{Offline Experiments} 
\begin{figure}
\centering  \subfloat[Subject S1 accuracy.]{\includegraphics[width=0.33\linewidth]{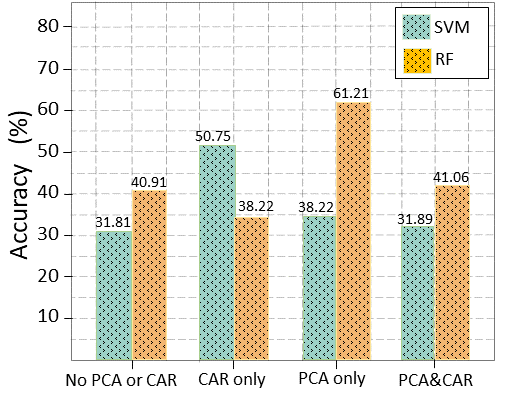}}
   \hfill
  \subfloat[Subject S2 accuracy.]{\includegraphics[width=0.33\linewidth]{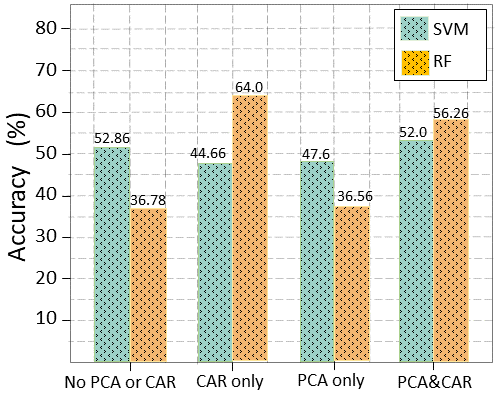}}
     \hfill
    \subfloat[Subject S3 accuracy.]{\includegraphics[width=0.33\linewidth]{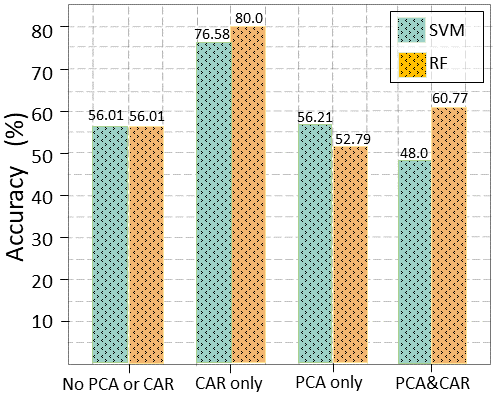}}
  \hfill
  \subfloat[Subject S1 ITR.]{\includegraphics[width=0.33\linewidth]{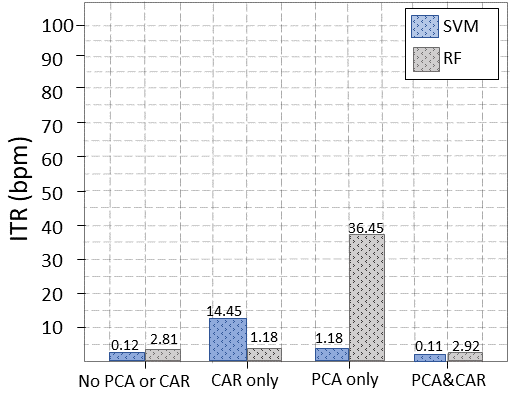}}
    \subfloat[Subject S2 ITR.]{\includegraphics[width=0.33\linewidth]{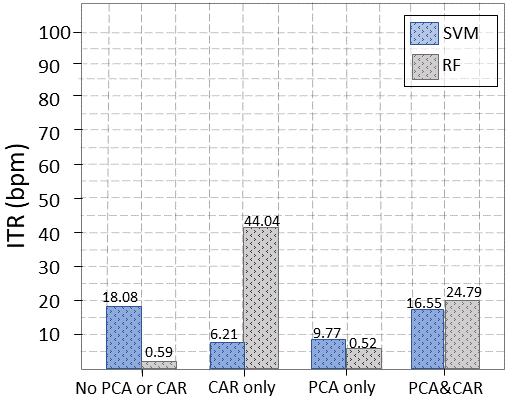}}
    \hfill
  \subfloat[Subject S3 ITR.]{\includegraphics[width=0.33\linewidth]{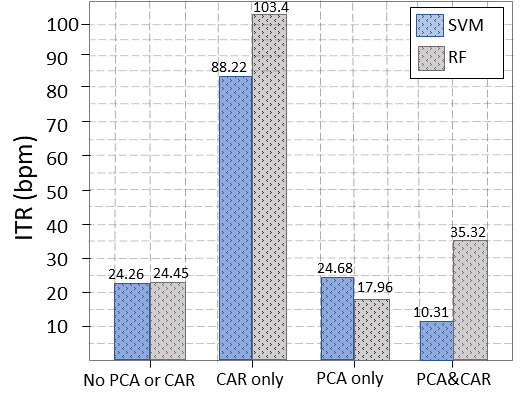}}
 
  \caption{Accuracy for offline experiments on all channels and PC.}
  \label{Images/offline_PC_AllC}
  \end{figure}

In order to examine the performance of the proposed system before integrating it with the Hololens AR set, we sought to design an experiment that runs on a PC first. The performance of using all channels in the analysis is examined in an offline setup in which subjects were instructed to focus on specific buttons (targets) in a pre-specified sequence. \fref{Images/offline_PC_AllC} shows the accuracy obtained for each subject using the PC. The figure compares the performance for different pre-processing stages (No CAR or PCA, using PCA only, using CAR only, and using CAR and PCA) and different classifiers (SVM and RF). The mean accuracy using all channels for the PC experiments reached 80\%.

Given that channels $O1$ and $O2$ are located on the visual cortex that is responsible for processing visual inputs \cite{chen2020stimulus}, we next examined the performance when signals from channels $O1$ and $O2$ only were used as opposed to using all channels. \fref{Images/offline_PC_two} demonstrates the accuracy achieved for the same subjects using different pre-processing stages and classifiers. The figure demonstrates an enhancement in the maximum classification accuracy for subjects S1 (69.33\% and 60.68 bits/min) and S2 (79.47\% and 100.98 bits/min) compared to using all channels in the analysis, while subject S3's maximum classification accuracy remained unchanged (80\% and 103 bits/min). This confirms our hypothesis that using $O1$ and $O2$ only would have a positive impact on the performance. 

\subsubsection{ Online Experiments}
\begin{figure}
  \centering
  \subfloat[Subject S1 accuracy.]{\includegraphics[width=0.33\linewidth]{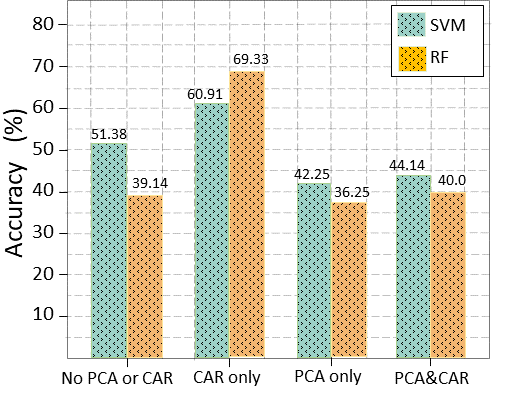}}
      \hfill
  \centering
  \subfloat[Subject S2 accuracy.]{\includegraphics[width=0.33\linewidth]{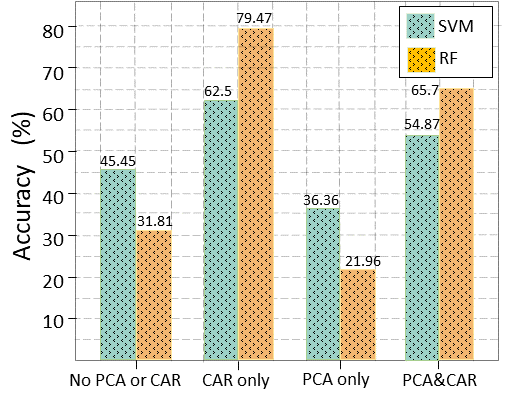}}
       \hfill
       \subfloat[Subject S3 accuracy.]{\includegraphics[width=0.33\linewidth]{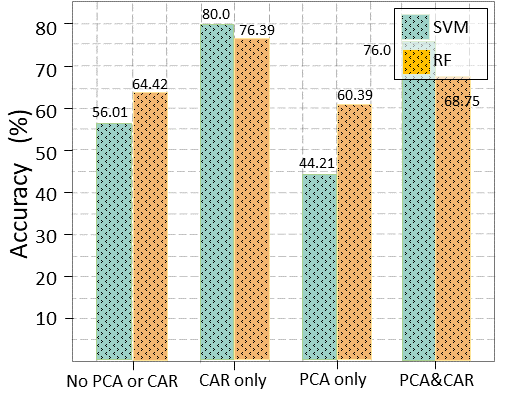}}
  \hfill
  \subfloat[Subject S1 ITR.]{\includegraphics[width=0.33\linewidth]{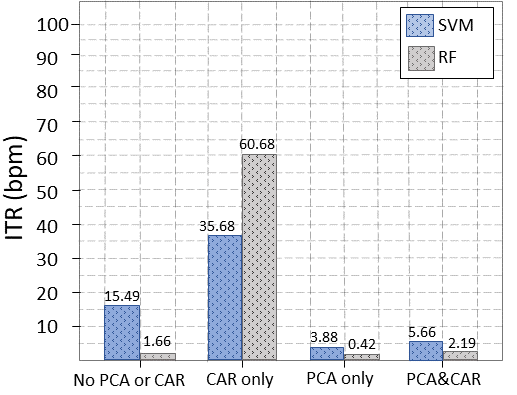}}
  \subfloat[Subject S2 ITR.]{\includegraphics[width=0.33\linewidth]{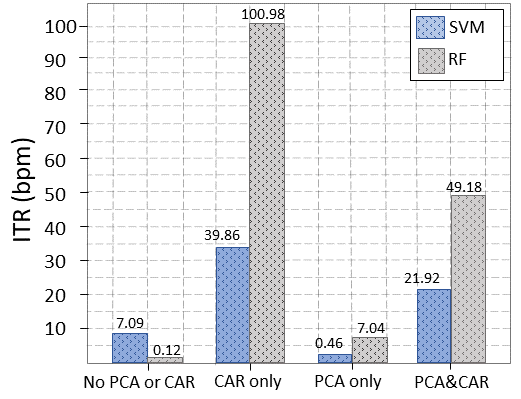}}
         \hfill
  \subfloat[Subject S3 ITR.]{\includegraphics[width=0.33\linewidth]{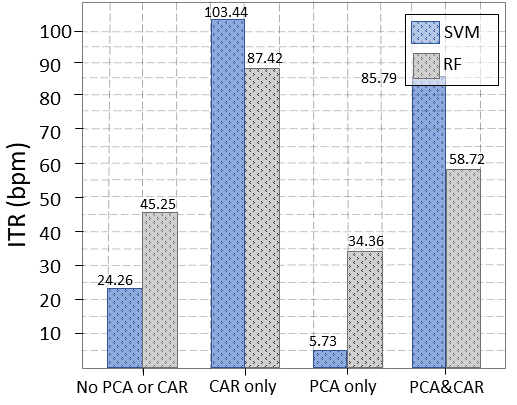}}
    
 \caption{Accuracy for offline experiments using $O1$ and $O2$ channels and PC.
 }
   \label{Images/offline_PC_two}
   \end{figure}

We next examined the feasibility of using the proposed approach in an online setup. The subject is then allowed to freely focus on any of the targets while the system attempts to identify the target intended. Given that the offline analysis indicates that using $O1$ and $O2$ only enhances the performance, in the online analysis, only signals from these two channels were considered. However,  as \fref{Images/offline_PC_two} demonstrates for the offline analysis, we faced a problem that there is not a single pre-processing or classification technique that outperforms all other techniques for all subjects. For Subject S1, the highest accuracy 69.33\% was obtained using CAR filter pre-processing and RF classifier. For Subject S3, the highest accuracy 80.0\% was obtained using CAR pre-processing and SVM classifier. This problem motivated the examination of the ensemble classification technique as it could improve the classification accuracy by voting across all pre-processing and classification techniques. Table \ref{tab:PC_Ensemble} summarizes the ensemble accuracy results obtained using the PC. The accuracies achieved are within the range of the accuracies achieved in the offline analysis, indicating that there is no overfitting. 
   
\begin{table}
\centering
\setlength{\tabcolsep}{1em}
\centering\arraybackslash
\renewcommand{\arraystretch}{1.3}
\caption{Accuracy in online experiments by the Ensemble classifier using channels $O1$ and $O2$ on PC.
}
\begin{tabular}{ccc} 
\hline
\textbf{Subject} & \textbf{Online Accuracy (\%)}  & \textbf{ITR (bpm)}   \\ 
\hline
S1      &   77.29  &     91.28       \\ 
\hline
S2        &  72.22   &          70.94        \\ 
\hline
S3       &   81.32     &     109.73      \\
\hline
\end{tabular}
\label{tab:PC_Ensemble}
\end{table}

\subsection{Experiments using HoloLens}
\subsubsection{Offline Experiments}
\begin{figure}
  \centering
  \subfloat[Subject S4 accuracy.]{\includegraphics[width=0.33\linewidth]{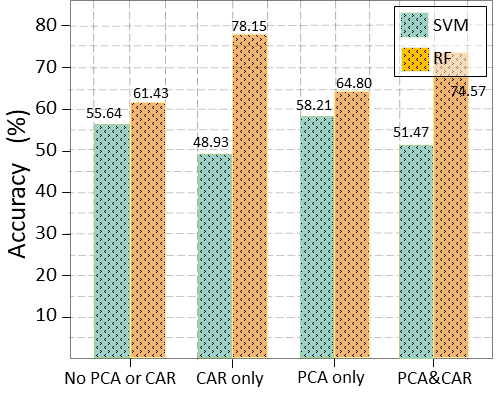}}
   \hfill
  \subfloat[Subject S5 accuracy.]{\includegraphics[width=0.33\linewidth]{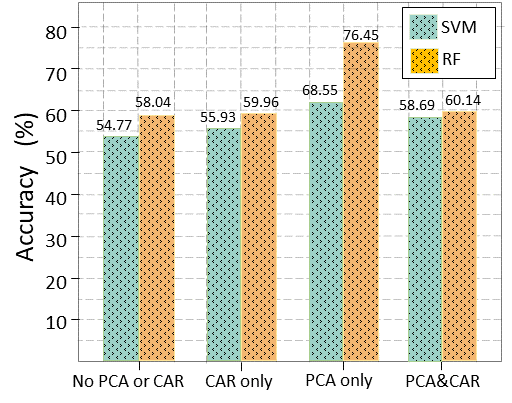}}
  \hfill
    \subfloat[Subject S6 accuracy.]{\includegraphics[width=0.33\linewidth]{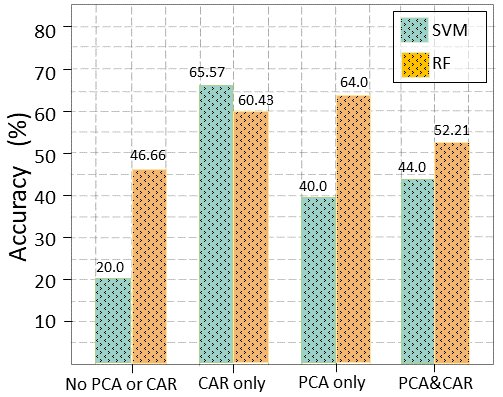}}
 \hfill
  \subfloat[Subject S4 ITR.]{\includegraphics[width=0.33\linewidth]{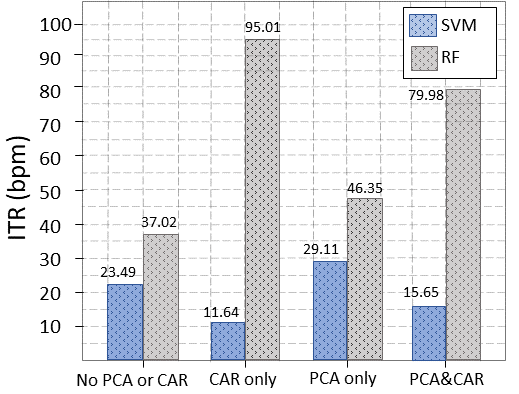}}
\hfill  
  \subfloat[Subject S5 ITR.]{\includegraphics[width=0.33\linewidth]{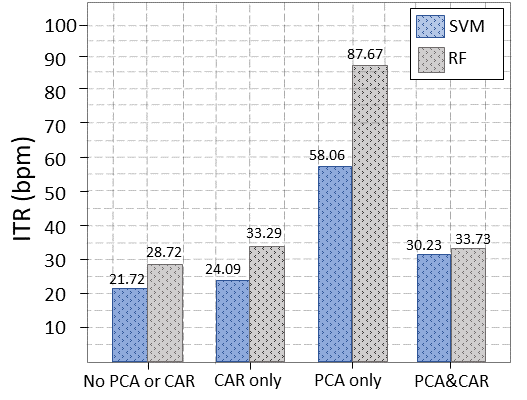}}
    \hfill
  \subfloat[Subject S6 ITR.]{\includegraphics[width=0.33\linewidth]{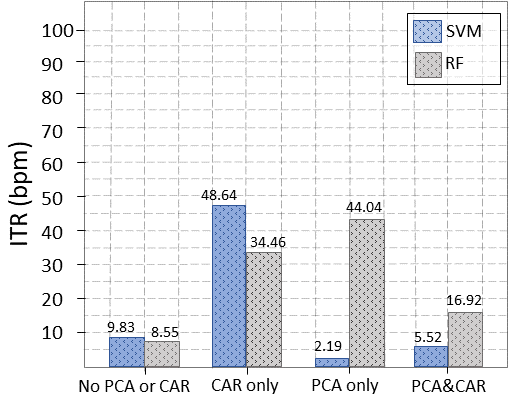}}
    
 \caption{Accuracy for offline experiments using $O1$ and $O2$ channels and Hololens.}
   \label{Images/offline_Hololens_two}
\end{figure}

We next examined the integrated system that uses the Hololens AR set, which is the main objective of this work. In these experiments, we used signals recorded from channels $O1$ and $O2$ only given the enhancement achieved using these channels in the PC experiments. We first examined the performance in an offline setup. \fref{Images/offline_Hololens_two} demonstrates the accuracy obtained for each of the subjects when testing on the HoloLens for different pre-processing stages and classifiers. The mean accuracy achieved across all three subjects using $O1$ and $O2$ channels on HoloLens is 73.39\%. Comparing these results to what was achieved using the PC indicates that using HoloLens is just as effective as using a PC as there is no significant difference between the accuracy of the HoloLens and the PC.
 
\subsubsection{Online Experiments}
Consistent with the PC experiments, the results in \fref{Images/offline_Hololens_two} emphasize that no certain combination of pre-processing and a classifier gives the best performance. For example, For Subject S4, the highest accuracy of 78.15\% (95.01 bits/min) was obtained using CAR filter pre-processing and Random Forests classifier. For Subject S5, the highest accuracy of 76.45\% (87.67 bits/min) was obtained using PCA pre-processing and Random Forests classifier. For Subject S6, the highest accuracy of 65.57\% (48.64 bits/min) was obtained using CAR filter pre-processing and SVM classifier. Therefore, the ensemble classification technique is used for the online experiments. Table \ref{tab:Hololens_Ensemble} summarizes the accuracies achieved in the HoloLens online Experiments. The results indicate a mean accuracy across all subjects of 76.2\%. To assess the significance of the results obtained from the ensemble classifier in comparison to the best-performing individual classifiers, a paired t-test was conducted. The analysis revealed a statistically significant difference, with a p-value of 0.00855, indicating that the performance of the ensemble classifier outperformed that of the individual classifiers.
\begin{table}
\setlength{\tabcolsep}{0.6em}
\centering
\renewcommand{\arraystretch}{1.12}
\caption{Accuracy in online experiments by the Ensemble classifier using channels $O1$ and $O2$ on Hololens.
}
\begin{tabular}{ccc} 
\hline
\textbf{Subject} & \textbf{Online Accuracy (\%)}  & \textbf{ITR (bpm)}  \\ 
\hline
S4      & 75.17        & 82.38     \\ 
\hline
S5      &  77.76   &    93.29         \\ 
\hline
S6      & 75.58        &   76.39       \\
\hline
\end{tabular}
\label{tab:Hololens_Ensemble}
\end{table}
\section{Conclusion}
We present a proof-of-concept study that shows the feasibility of using SSVEP-based BCI input with allowed head rotations to navigate through the AR environment without conventional BCI-imposed restrictions such as staying still. Head movements are not only allowed between SSVEP commands but also during the SSVEP commands. No significant influence on the results was observed despite the movements, and the results are comparable to prior work such as \cite{milsap2021low,si2018towards} whose performance dropped significantly when movements occurred. In addition, the required stimulation time of this study is only 5 seconds which is less than the stimulation time needed by previous studies \cite{milsap2021low,yue2021exploring,si2018towards}. A paired t-test revealed a significant difference between the results obtained from our ensemble classifier and the best-performing individual classifier. Online and offline experiments were done with subjects of different genders and hair types. Using channels $O1$ and $O2$ only instead of using all 14 channels significantly increased the mean accuracy.  Then, an ensemble classifier is used to overcome the inconsistent performance among subjects. Moreover, our BCI-AR integration study shows that on a simple HoloLens, we can achieve accuracy comparable to on a PC and provide seamless communication in an AR environment.

%
%
%
\bibliographystyle{ieeetran}
\bibliography{mybibliography}

\end{document}